\documentclass[english,twoside,a4paper,10pt]{article}

\usepackage[latin1]{inputenc}
\usepackage[T1]{fontenc}
\usepackage{amsmath}
\usepackage{amsfonts}
\usepackage{graphicx}
\usepackage{subfigure}
\usepackage{a4wide}
\usepackage{amssymb}
\usepackage{fancyhdr}
\usepackage{mathrsfs}
\usepackage[toc,page]{appendix}

\begin{document}

\title{\bf Chaotic inflation in higher derivative gravity theories}
\author{ 
Shynaray Myrzakul\footnote{Email: shynaray1981@gmail.com},\,\,\,
Ratbay Myrzakulov\footnote{Email: rmyrzakulov@gmail.com},\,\,\,
Lorenzo Sebastiani\footnote{E-mail address: l.sebastiani@science.unitn.it
}\\
\\
\begin{small}
Department of General \& Theoretical Physics and Eurasian Center for
\end{small}\\
\begin{small} 
Theoretical Physics, Eurasian National University, Astana 010008, Kazakhstan
\end{small}\\
}

\date{}

\maketitle


\begin{abstract}
In this paper, we investigate chaotic inflation from scalar field subjected to potential in the framework of $f(R^2, P, Q)$-gravity, where we add a correction to Einstein's gravity based on a function of the square of the Ricci scalar $R^2$, the contraction of the Ricci tensor $P$, and the contraction of the Riemann tensor $Q$. The Gauss-Bonnet case is also discussed. We give the general formalism of inflation, deriving the slow-roll parameters, the $e$-folds number, and the spectral indexes. Several explicit examples are furnished, namely we will consider the cases of massive scalar field and scalar field with quartic potential and some power-law function of the curvature invariants under investigation in the gravitational action of the theory. Viable inflation according with observations is analyzed. 
\end{abstract}



\tableofcontents
\section{Introduction}

The last data \cite{WMAP, Planck} coming from observations of the universe anisotropy increased the interest for inflationary universe. 
Inflation has been proposed several years ago~\cite{Guth, Sato} to solve the problems of the initial conditions
of Friedmann universe and eventually to explain some
issue related to the particle physics, but,
despite to the constraints that must be satisfied to fit the cosmological data,
the choice of the models is quite large  (see Refs.~\cite{rev1, rev2} for an introduction to  inflationary cosmology).

Many inflationary models are based on scalar field representation, where
an homogeneous scalar field, called  inflaton, is subjected to a potential and produces the accelerated expansion of early-time universe, when the curvature is near to the Planck scale.
Typically, the magnitude of the inflaton is arbitrarily large (chaotic inflation) at the beginning of the 
inflation~\cite{chaotic}  and therefore
the field rolls down towards a potential minimum where acceleration ends: thus, the field starts to oscillate and the rehating processes for the particle production take place
~\cite{buca1, buca2, buca3, buca4}. Other inflationary models are based on a phase transition between two scalar 
fields~\cite{ibrida1, 
ibrida2}.
Additionally, it is expected that inflation 
is related with quantum corrections to General Relativity, and in this direction many efforts to construct viable models taking into account higher derivative corrections to General Relativity emerging at the Planck scale have been done~\cite{q1, q1bis, q2, q3}.

In this work, we would like to investigate how chaotic inflation works in the framework of higher derivative corrections to the theory of Einstein. Our modification to the gravitational action of General Relativity  is expressed in terms of the square of the Ricci scalar and the contractions of Ricci and Riemann tensors, in the attempt to include a wide class of models (see Refs.~\cite{mod1,mod2, mod3} for reviews). We will present some explicit examples of theories and potentials for inflaton which make viable the inflation according with cosmological data. 

The paper is organized in the following way. In Chapter {\bf 2}, we present the model:
the Hilbert-Einstein action of General Realativity is modified by adding a function of the square of the Ricci scalar and the contractions of Ricci and Riemann tensors, and the contribute of a scalar field subjected to a potential is also included. Thus,
we
derive the Lagrangian and the related Equations of motion of the theory in flat Friedmann-Robertson-Walker space-time, by using a method based on Lagrangian multipliers which reduces the Equations of motion at the second order. It is interesting to note how for Friedmann-Robertson-Walker metric such a kind of models has an equivalent description as Gauss-Bonnet theory. In Chapter {\bf 3}, we investigate the general feauture of chaotic inflation from scalar field in the framework of our modified theory. Chapter {\bf 4} and Chapter {\bf 5} are devoted to the explicit examples, corresponding to two well-know cases of chaotic inflation, namely chaotic inflation from massive scalar field and chaotic inflation from field with quartic potential. In the framework of General Relativity they lead to viable inflation, and we are interested to see how results change for some toy model of Gauss-Bonnet modified gravity and some model based on the power-law functions of the curvature invariants under investigation. The conclusions with some final remarks are given in Chapter {\bf 5}.

We use units of $k_{\mathrm{B}} = c = \hbar = 1$ and denote the
gravitational constant, $G_N$, by $\kappa^2\equiv 8 \pi G_{N}$, such that
$G_{N}=1/M_{\mathrm{Pl}}^2$, $M_{\mathrm{Pl}} =1.2 \times 10^{19}$ GeV being the Planck mass.


\section{Formalism}

Let us consider the following action,
\begin{equation}
I=\int_\mathcal{M}\sqrt{-g} \left[\frac{R}{2\kappa^2}+f(R^2,P,Q)-\frac{1}{2}g^{\mu\nu}\partial_{\mu}\phi\partial_{\nu}\phi-V(\phi)\right]\,,
\quad \kappa^2=\frac{8\pi}{M_{Pl}^2}\,,\label{action}
\end{equation}
where $\mathcal{M}$ is the space-time manifold, $g$ is the determinant of the metric tensor $g_{\mu\nu}$ and $R$ is the Ricci scalar. The gravitational part of the Lagrangian takes into account the higher derivative corrections to Einstein's gravity encoded in the generic function $f(R^2, P, Q)\equiv f$ where
\begin{equation}
P=R_{\mu\nu}R^{\mu\nu}\,,\quad Q=R_{\mu\nu\sigma\xi}R^{\mu\nu\sigma\xi}\,,
\end{equation}
$R_{\mu\nu}$ and $R_{\mu\nu\sigma\xi}$ being the Ricci tensor and the Riemann tensor, respectively. The ``matter'' part of the Lagrangian depends on a scalar field $\phi$ subjected to the potential $V(\phi)$.

We will work with the flat Friedmann-Robertson-Walker (FRW) metric, whose general expression is given by
\begin{equation}
d s^2=-N(t)^2 dt^2+a(t)^2(dx^2+dy^2+dz^2)\,,\quad\sqrt{-g}=N(t)a(t)^3\,,\label{metric}
\end{equation}
where $N(t)\equiv N$ is a lapse function and $a(t)\equiv a$ is the scale factor, the both depending on the cosmological time $t$. Thus, the curvature invariants of the model on flat FRW space-time read
\begin{equation}
R=\frac{6}{N^2}\left(X+Y\right)\,,\quad P=\frac{12}{N^2}\left(X^2+Y^2+XY\right)\,,\quad Q=\frac{12}{N^2}\left(X^2+Y^2\right)\,,\label{RPQ}
\end{equation}
with
\begin{equation}
X=\frac{\ddot a}{a}-\frac{\dot a}{a}\frac{\dot N}{N}\,,\quad Y=\frac{\dot a^2}{a^2}\,,
\end{equation}
and the dot denotes the derivative with respect to the time.
By plugging this expressions into the action (\ref{action}), we obtain an higher derivative Lagrangian. However, by using a method based on the Lagrangian multipleyer~\cite{L1, L2, Monica, miolag}, we can deal with a first order standard Lagrangian.
We introduce the Lagrangian multipliers $\zeta\,,\sigma\,,\xi$
as
\begin{eqnarray}
I&=&\int_\mathcal{M} Na^3\left[\frac{3}{\kappa^2 N^2}\left(X+Y\right)+f(R^2,P,Q)
-\zeta\left(R-\frac{6}{N^2}\left(X+Y\right)\right)\right.\nonumber\\
&&\left.
-\sigma\left(P-\frac{12}{N^4}\left(X^2+Y^2+XY\right)\right)
-\xi\left(Q-\frac{12}{N^4}\left(X^2+Y^2\right)\right)
+\frac{1}{2N^2}\dot\phi^2-V(\phi)\right]\,.
\end{eqnarray}
In order to get a first order Lagrangian, we rewrite the action as 
\begin{eqnarray}
I&=&\int_\mathcal{M} Na^3\left[\frac{3}{\kappa^2 N^2}\left(X+Y\right)+f(R^2,P,Q)
-\zeta\left(R-\frac{6}{N^2}\left(X+Y\right)\right)\right.\nonumber\\
&&\left.
-\sigma\left(P-\frac{R^2}{3}\right)
-\xi\left(Q-\frac{R^2}{3}\right)
-\frac{12}{N^4}XY(\sigma+2\xi)
+\frac{1}{2N^2}\dot\phi^2-V(\phi)\right]\,,
\end{eqnarray}
such that the variations with respect to $R\,,P$ and $Q$ lead to
\begin{equation}
\zeta=f_{R}(R^2, P, Q)+\frac{2R}{3}\left[f_{P}(R^2, P, Q)+f_{Q}(R^2,P,Q)\right]\,,\quad\sigma=f_{P}(R^2, P, Q)\,,\quad\xi=f_{Q}(R^2, P, Q)\,,\label{const}
\end{equation}
where $f_{R, P, Q}(R^2, P, Q)$ are the derivatives of $f(R^2, P, Q)$ respect to $R\,,P$ and $Q$, 
\begin{equation}
f_R(R^2, P, Q)\equiv 2R\frac{d f(R^2, P, Q)}{d R^2}\,,\quad
f_P(R^2, P, Q)\equiv \frac{d f(R^2, P, Q)}{d P}\,,\quad
f_Q(R^2, P, Q)\equiv \frac{d f(R^2, P, Q)}{d Q}\,.
\end{equation}
Thus,
after integration by part, we obtain for the gravitational part
\begin{eqnarray}
\mathcal L_{grav}(a,\dot a, N, R, \dot R, P, \dot P, Q, \dot Q)&=&
-\frac{3 a \dot a^2}{\kappa^2 N}+f N a^3-\zeta R Na^3-\frac{6a\dot a^2\zeta}{N}-\frac{6\dot a a^2\dot \zeta}{N}
\nonumber\\&&\hspace{-1cm}
-\sigma\left(P-\frac{R^2}{3}\right)N a^3
-\xi\left(Q-\frac{R^2}{3}\right)N a^3
+\frac{4\dot a^3}{N^3}\left(\dot\sigma+2\dot\xi\right)\,,\label{gL}
\end{eqnarray}
and the total Lagrangian results to be
\begin{equation}
\mathcal L_{tot}=\mathcal L_{grav}+\mathcal L_{\phi}\,,\quad \mathcal L_{\phi}=\left(\frac{1}{2N}\dot\phi^2-V(\phi)N\right) a^3\,.\label{totalL}
\end{equation}
As a result, we obtained a first order Lagrangian respect to the unknown variables $N(t)\,,a(t)\,, R(t)\equiv R\,,P(t)\equiv P\,,Q(t)\equiv Q$.

Some remarks are in order. Obviously, the expression (\ref{gL}) can be generalized to the case $f(R^2,P,Q)\rightarrow f(R, P, Q)$. An interesting special case is given by Gauss-Bonnet gravity. The Gauss-Bonnet four dimensional topological invariant reads
\begin{equation}
G=R^2-4P+Q\,,\quad G=\frac{24}{N^4}XY\,,\label{Ggeneric}
\end{equation}
where the second expression is the form of the Gauss-Bonnet on the flat FRW space-time. If $f(R, P, Q)=f(R, G)$, by taking into account that
\begin{equation}
f _R (R, P, Q)\rightarrow f_R(R, G)+2Rf_G(R,G)\,,\quad
f _P (R, P, Q)\rightarrow -4 f_G(R, G)\,,\quad
f _Q (R, P, Q)\rightarrow f_G(R, G)\,,\label{GG}
\end{equation}
we derive from (\ref{gL}),
\begin{eqnarray}
\mathcal L_{grav}(a,\dot a, N, R, \dot R, G, \dot G)=
-\frac{3 a \dot a^2}{\kappa^2 N}+N a^3(f-R f_R- G f_G)-\frac{6a\dot a^2 f_R}{N}-\frac{6\dot a a^2\dot f_R}{N}
-\frac{8\dot a^3}{N^3}\dot f_G\,,\label{GL}
\end{eqnarray}
according with Ref.~\cite{Monica}. Moreover, it is possible to demonstrate that the Lagrangian of $f(R,P,Q)$-models corresponds to the Lagrangian of $f(R,G)$-theories on FRW background~\cite{M1,M2}. We may replace $Q$ with
$Q=G-R^2+4P$ and therefore $f_Q(R,P,Q)$ with $f_G(R,G,Q)$ in (\ref{gL}) and make the following substitutions
\begin{equation}
f_R(R,P,Q)\rightarrow f_R(R,G,Q)+2R f_G(R,G,Q)\,,\quad
f_P(R,P,Q)\rightarrow f_P(R,G,Q)-4 f_G(R,G,Q)\,,
\end{equation}
in order to cancel the additional derivatives that we have aquired. We get
\begin{eqnarray}
\mathcal L_{grav}(a,\dot a, N, R, \dot R, G, \dot G, Q, \dot Q)&=&
-\frac{3 a \dot a^2}{\kappa^2 N}+N a^3(f-R f_R- G f_G)-\frac{6a\dot a^2 f_R}{N}-\frac{6\dot a a^2\dot f_R}{N}
-\frac{8\dot a^3}{N^3}\dot f_G\nonumber\\
&&\hspace{-1cm}-f_P\left(P+Na^3 \frac{R^2}{3}+\frac{4a\dot a^2 R}{N}+\frac{4a^2\dot a \dot R}{N}\right)+
\dot f_P\left(\frac{4\dot a^3}{N^3}-\frac{4a^2\dot a R}{N}\right)\,,
\end{eqnarray}
and after integration by part we obtain
\begin{eqnarray}
\mathcal L_{grav}(a,\dot a, N, R, \dot R, G, \dot G, Q, \dot Q)&=&
-\frac{3 a \dot a^2}{\kappa^2 N}+N a^3(f-R f_R- G f_G)-\frac{6a\dot a^2 f_R}{N}-\frac{6\dot a a^2\dot f_R}{N}
-\frac{8\dot a^3}{N^3}\dot f_G\nonumber\\
&&\hspace{-1cm}+f_P Na^3\left(\frac{R^2}{3}-\frac{G}{2}-P\right)\,,
\end{eqnarray}
but on FRW metric the last term is null and we recover (\ref{GL}).
To pass from $f(R^2, P, Q)$ to $f(R, G)$-gravity on FRW space-time, we can substitute $R^2\,,P\,,Q$ with $R^2\,,G\,,C^2$ into the action (\ref{action}). Here, $C^2$ is the ``square'' of the Weyl tensor,
\begin{equation}
C^2=\frac{1}{3}R^2-2R_{\mu\nu}R^{\mu\nu}+R_{\xi\sigma\mu\nu}R^{\xi\sigma\mu\nu}\,,
\end{equation}
and the following relations are met,
\begin{equation}
P=\frac{C^2}{2}-\frac{G}{2}+\frac{R^2}{3}\,,\quad
Q=2C^2-G+\frac{R^2}{3}\,.\label{superrelation}
\end{equation}
On FRW metric (\ref{metric}), the square of the Weyl tensor is identically null ($C^2=0, \delta C^2=0$) and does not contribute to the dynamics of the model, such that we can drop down it from the Lagrangian and use the formalism of $f(R^2, G)$-gravity.\\
\\
By making the variation of (\ref{gL})--(\ref{totalL}) respect to $N(t)$ and therefore by putting $N(t)=1$, we find
\begin{equation}
\frac{3H^2}{\kappa^2}+\left(f-\zeta R\right)+6H^2\zeta+6H\dot\zeta-\sigma\left(P-\frac{R^2}{3}\right)-\xi\left(Q-\frac{R^2}{3}\right)-12H^3\left(\dot\sigma+2\dot\xi\right)=\frac{\dot\phi^2}{2}+V(\phi)\,,\label{N}
\end{equation}
where $H=\dot a/a$ is the Hubble parameter. The variation respect to $a(t)$  with $N(t)=1$ leads to
\begin{eqnarray}
&&\frac{1}{\kappa^2}\left(3H^2+2\dot H\right)+\left(f-\zeta R\right)+\left(4\dot H \zeta+6H^2\zeta\right)+4H\dot\zeta+2\ddot\zeta-\sigma\left(P-\frac{R^2}{3}\right)
-\xi\left(Q-\frac{R^2}{3}\right)
\nonumber\\&&
-8H\left(H^2+\dot H\right)
\left(\dot\sigma+2\dot\xi\right)-4H^2\left(\ddot\sigma+2\ddot\xi\right)=-\left(\frac{\dot\phi^2}{2}-V(\phi)\right)\,.\label{a}
\end{eqnarray}
Finally, the variations respect to $R\,,P$ and $Q$ for the gauge $N(t)=1$ read
\begin{equation}
R=12H^2+6\dot H\,,\quad P=12\left(\dot H^2+3H^4+3 \dot H H^2 \right)\,,\quad
Q=12\left(\dot H^2+2H^4+2 \dot H H^2 \right)\,.\label{RPQ2}
\end{equation}
In conclusion, we obtained a set of five second order differential equations (\ref{N})--(\ref{RPQ2}). By taking the time derivative of (\ref{N}) and therefore by using (\ref{a}) we derive the energy conservation law of the field,
\begin{equation}
\ddot\phi+3H\dot\phi=-V'(\phi)\,.\label{conslaw}
\end{equation}
Here, the prime denotes the derivative with respect to $\phi$. Thus, we can make the following identification,
\begin{equation}
\rho_\phi=\frac{\dot\phi^2}{2}+V(\phi)\,,\quad p_\phi=\frac{\dot\phi^2}{2}-V(\phi)\,,
\quad \dot\rho_\phi+3H(\rho_\phi+p_\phi)\,,
\end{equation}
where $\rho_\phi$ and $p_\phi$ are the effective energy density and pressure of the field, respectively. We also may introduce the Equation of State (EoS) parameter of the field as
\begin{equation}
\omega_\phi\equiv\frac{p_\phi}{\rho_\phi}=\frac{\dot\phi-2V(\phi)}{\dot\phi+2V(\phi)}\,.\label{omega}
\end{equation}
Let us see now how the inflationary cosmology is reproduced by such a kind of models.

\section{Inflationary cosmology}

We would like to see how a modification to Einstein's gravity based on the higher derivative correction terms $R^2\,,P\,,Q$ changes the classical picture of the inflation from scalar field models in the General Relativity background.
The dynamics of the model (\ref{action}) is governed by the equations (\ref{N}) and (\ref{conslaw}) with (\ref{RPQ2}). 

The inflation is described by a quasi-de Sitter solution, with the Hubble parameter which slowly decreases with the time, such that the slow-roll approximations are valid, 
\begin{equation}
|\frac{\dot H}{H^2}|\ll 1\,,\quad |\frac{\ddot H}{H\dot H}|\ll
1\,.\label{slowrollcondition}
\end{equation}
It means, that the magnitude of the so called ``slow-roll parameters'',
\begin{equation}
\epsilon=-\frac{\dot H}{H^2}\,,\quad\eta=-\frac{\dot H}{H^2}-\frac{\ddot H}{2 H\dot
H}\equiv2\epsilon-\frac{1}{2\epsilon H}\dot\epsilon\,,\label{slowrollpar}
\end{equation}
must be small during inflation. Moreover, $0<\epsilon$ in order to have $\dot H<0$, and, since the acceleration is expressed as
\begin{equation}
\frac{\ddot a}{a}=\dot H+H^2\,,
\end{equation}
we see that the accelerated expansion finishes only when the $\epsilon$ slow-roll parameter is on the order of the unit.

To describe the early-time acceleration, we will use the approach of chaotic inflation. At the beginning the field, namely the inflaton, is assumed to be negative and very large. The slow roll  regime takes place if the kinetic energy of the field is much smaller respect to the potential,
\begin{equation}
\dot\phi^2\ll V(\phi)\,,\quad |\ddot\phi|\ll 3H \dot\phi\label{phiV}
\end{equation}
In this way, the field EoS parameter (\ref{omega}) is $\omega_\phi\simeq-1$ and the de Sitter solution can be realized.
On the other hand, from (\ref{conslaw})  we have, in the slow-roll approximation (\ref{phiV}), 
\begin{equation}
3H\dot\phi\simeq -V'(\phi)\,,\quad 3H\ddot\phi\simeq -V''(\phi)\dot\phi\,,\label{cons2}
\end{equation}
and, if
$V'(\sigma)>0$, the kinetic energy increases with the field which tends to a minimum of the potential, where (\ref{phiV}) is not valid and inflation ends.\\
\\
For the (quasi) de Sitter solution of inflation, one introduces the $e$-folds number as
\begin{equation}
\mathcal N\equiv\log\left[\frac{a_\text{f}}{a_\text{i}}\right]
=\int^{t_\text{f}}_{t_\text{i}} H(t)dt\simeq 3\int^{\phi_\text{i}}_{\phi_\text{f}} \frac{H^2}{V'(\phi)}d\phi
\,,\label{Nfold}
\end{equation}
where $a_\text{i,f}$ are the scale factor at the beginning and at the end of inflation,  $t_\text{i,f}$ the related times and $\phi_\text{i,f}$ the values of the field at the beginning and at the end of inflation.
Generally, the primordial acceleration can solve the horizon and velocities problems of Friedmann universe
if $55<\mathcal N$.

By using the slow-roll parameters, one can also evaluate the universe anisotropy coming from inflation deriving the
spectral indexes.
The amplitude of the primordial scalar power spectrum reads
\begin{equation}
\Delta_{\mathcal R}^2=\frac{\kappa^2 H^2}{8\pi^2\epsilon}\,,\label{spectrum}
\end{equation}
and according with cosmological observations must be $\Delta_{\mathcal R}^2\simeq 10^{-9}$;
the spectral index $n_s$ and the tensor-to-scalar 
ratio $r$ are given by
\begin{equation}
n_s=1-6\epsilon+2\eta\,,\quad r=16\epsilon\,.\label{index}
\end{equation}
The last results observed by the Planck satellite~\cite{Planck} constrain these quantities as
$n_{\mathrm{s}} = 0.9603 \pm 0.0073\, (68\%\,\mathrm{CL})$ and 
$r < 0.11\, (95\%\,\mathrm{CL})$.

For chaotic inflation in the background of General Relativity, namely in the case of action (\ref{action}) with $f(R^2\,,P\,,Q)=0$, one has in terms of the field potential and its derivative,
\begin{equation}
\epsilon=\frac{1}{2\kappa^2}\left(\frac{V'(\phi)}{V(\phi)}\right)^2\,,\quad 
\eta=\frac{1}{\kappa^2}\left(\frac{V''(\phi)}{V(\phi)}\right)\,,\quad
\mathcal N=\kappa^2\int^{\phi_i}_{\phi_e}\frac{V(\phi)}{V'(\phi)}d\phi\,,\label{ciao}
\end{equation}
where the quasi de Sitter solution of inflation is given by $H_\text{dS}^2=\kappa^2 V(\phi)/3$, while,
in the slow-roll approximation, $\dot H\simeq \kappa^2 V'(\phi)\dot\phi/(6H_\text{dS})$ with $\dot\phi$ derived from  (\ref{cons2}).

In our case, the quasi de Sitter solution of inflation $H\simeq H_\text{dS}$, $H_\text{dS}$ being a constant, is given by equation (\ref{N}) under the condition (\ref{phiV}), namely
\begin{eqnarray}
&&\frac{3H^2_\text{dS}}{\kappa^2}+\left(f(R^2_\text{dS}, P_\text{dS}, Q_\text{dS})-\frac{R_\text{dS}f_R(R^2_\text{dS}, P_\text{dS}, Q_\text{dS})}{2}\right)-\frac{R_\text{dS}^2f_Q(R^2_\text{dS}, P_\text{dS}, Q_\text{dS})}{6}
\nonumber\\
&&
-\frac{R_\text{dS}^2f_P(R^2_\text{dS}, P_\text{dS}, Q_\text{dS})}{4}=V(\phi)\,,\label{equazioneds}
\end{eqnarray}
with
\begin{equation}
R_\text{dS}=12 H_\text{dS}^2\,,\quad
P_\text{dS}=\frac{R_\text{dS}^2}{4}\,,\quad Q_\text{dS}=\frac{R_\text{dS}^2}{6}\,.
\end{equation}
By taking the derivative of (\ref{N}) with (\ref{RPQ2}) and by using the slow-roll conditions (\ref{slowrollcondition}) and (\ref{phiV}), we obtain the equations for $\dot H$, $\ddot H$,
\begin{eqnarray}
&&\hspace{-2cm}\dot H\left[\frac{6H}{\kappa^2}+12 H f_R-144 H^3 f_{RR}-5184 H^7 f_{PP}
-2304 H^7 f_{QQ}-1728 H^5 f_{RP}-1152 H^5 f_{RQ}-6912 H^7 f_{PQ}\right]\nonumber\\
&&
\simeq V'(\phi)\dot\phi\,,\label{Hdotgeneric}
\end{eqnarray}
\begin{eqnarray}
&&\hspace{-2cm}\ddot H\left[\frac{6H}{\kappa^2}+12 H f_R-144 H^3 f_{RR}-5184 H^7 f_{PP}
-2304 H^7 f_{QQ}-1728 H^5 f_{RP}-1152 H^5 f_{RQ}-6912 H^7 f_{PQ}\right]\nonumber\\
&&
\simeq 2V'(\phi)\ddot\phi-2HV'(\phi)\dot\phi\,\epsilon\,,\quad\epsilon=-\frac{\dot H}{H^2}\,,\label{Hdotdotgeneric}
\end{eqnarray}
where $H=H_\text{dS}$ and $\dot\phi$ is determined by (\ref{cons2}).  The last term of (\ref{Hdotdotgeneric}) has been approximated to simplify it: it comes from terms proportional to $\sim \dot H^2$, which are negligible if potential is flat (in such a case, $\epsilon\ll |\eta|$), but not in other cases like for the power-law potentials that we will analyze. Thus,
the dependences on $V(\phi), V'(\phi)$ and $V''(\phi)$ of the slow-roll parameters $\epsilon\,,\eta$ and of the $e$-folds number change respect to inflation in Einstein's background case. 

In the following chapters, we will consider some suitable potentials for scalar field and we will see how the early-time acceleration is reproduced according with cosmological data in $f(R^2, P, Q)$-gravity.

\section{Quadratic potential}

One classical example of chaotic inflation is given by massive field with potential
\begin{equation}
V(\phi)=\frac{m^2\phi^2}{2}\,,\quad 0<m\,,\label{pot1}
\end{equation}
where $m$ is a positive mass term smaller than the Planck mass, $m\ll M_{Pl}$, in order to avoid quantistic effects during inflation. Let us assume $\phi$ negative and very large. In the slow-roll regime (\ref{phiV}), equation~(\ref{cons2}) leads to
\begin{equation}
\log\left[\frac{\phi}{\phi_\text{i}}\right]=\frac{m^2}{3H_\text{dS}}(t-t_\text{i})+\text{const}\,,
\end{equation}
where $H_\text{dS}$ is the quasi-de Sitter solution of inflation given by (\ref{equazioneds}) and $\phi_\text{i}$ the value of the field at the beginning of inflation when $t=t_\text{i}$. For example, in Einstein's gravity with $f(R^2,P,Q)=0$, one has
\begin{equation}
H_\text{dS}=-\sqrt{\frac{\kappa^2}{6}}m\phi_\text{i}\equiv-2\sqrt{\frac{\pi}{3}}\frac{m\phi_\text{i}}{M_\text{Pl}}\,,\quad
\phi\simeq\phi_\text{i}+m\sqrt{\frac{2}{3\kappa^2}}(t-t_\text{i})\equiv\phi_\text{i}+\frac{m M_{Pl}}{\sqrt{12\pi}}(t-t_\text{i})\,.\label{38}
\end{equation}
Thus, the slow-roll approximation (\ref{phiV}) is valid as soon as
\begin{equation}
\frac{M_{Pl}}{\sqrt{12\pi}}<|\phi_\text{i}|<\frac{M_{Pl}^2}{2m}\sqrt{\frac{3}{\pi}}\,,\label{39}
\end{equation}
where we have also taken into account that $H_\text{dS}<M_{Pl}$: however, the accelerated expansion ends only when the $\epsilon$ slow-roll parameter is equal to one. Note 
that the field is larger than the Planck Mass, but its kinetic energy is smaller. 
The $e$-folds and the slow-roll parameters (\ref{ciao}) read
\begin{equation}
\epsilon\simeq\frac{M_{Pl}^2}{4\pi\phi_\text{i}^2}\,,\quad 
\eta\simeq\frac{M_{Pl}^2}{4\pi\phi_\text{i}^2}\,,\quad
\mathcal N\simeq\frac{2\pi\phi_\text{i}^2}{M_{Pl}^2}\,,\label{40}
\end{equation}
and for large $e$-folds (i.e. $\phi_\text{i}$ much larger than the Planck Mass) the slow-roll parameters 
are small and the spectral index $n_s$ in (\ref{index}) can satisfy the Planck data. However, we must stress that the tensor-to-scalar ratio $r$ results to be slightly bigger than the Planck bound. In general, this is true for all the power-law potential in scalar field representation (except for the case $V(\phi)\sim \phi$, which is negative for large values of the field and presents some criticism). In the last months, the correct value of the tensor-to-scalar ratio has been a debated question, and for this reason the analysis of such a kind of models is still interesting. For example, the last BICEP2 results~\cite{BICEP} 
indicated for the $B$-mode polarization of the CMB-radiation the tensor-to-scalar ratio 
$r =0.20_{-0.05}^{+0.07}\, (68\%\,\mathrm{CL})$,  
and the vanishing of $r$  
has been rejected at $7.0 \sigma$ level, while, as we stated before, the data coming from Planck experiment reveal an upper bound for the tensor-to-scalar ratio at $r=0.11$ with $95\%$ CL. Moreover, it should important to mention that very recent combinations of  the Planck and revised
BICEP2/Keck Array likelihoods lead to $r< 0.09$ with $95\%$~\cite{lll}.

Let us see now how inflation induced by massive scalar field works in some toy models of $f(R^2,P, Q)$-gravity. First of all, we will consider the subclass of Gauss-Bonnet gravity, and then we will extend our investigation to more general theories.

\subsection{Gauss-Bonnet models \label{GB1}}

The Gauss-Bonnet (\ref{Ggeneric}) in the flat FRW space-time (\ref{metric}) with the gauge $N(t)=1$ is given by
\begin{equation}
G=24H^2\left(H^2+\dot H\right)\,.
\end{equation}
For the case $f(R^2, P, Q)\equiv f(G)$, by taking into account (\ref{GG}), equation (\ref{N}) reads
\begin{equation}
\frac{3H^2}{\kappa^2}+(f-G f_G)+24H^3\dot G f_{GG}=\frac{\dot\phi^2}{2}+V(\phi)\,,\label{Gcase}
\end{equation}
where we have used the fact $\dot f_G=\dot G f_{GG}$.
Thus, the de Sitter solution is the corresponding of (\ref{equazioneds}),
\begin{equation}
\frac{3H_\text{dS}^2}{\kappa^2}+(f(G_\text{dS})-G_\text{dS} f_G(G_\text{dS}))=V(\phi)
\,,\quad
G_\text{dS}=24 H_\text{dS}^4\,,\label{45}
\end{equation}
and by taking the derivative of (\ref{Gcase}) one has in the slow-roll approximation (\ref{slowrollcondition}),
\begin{equation}
\frac{6H\dot H}{\kappa^2}-2304\dot H H^7 f_{GG}\simeq V'(\phi)\dot\phi\,,\label{46}
\end{equation}
which corresponds to (\ref{Hdotgeneric}) with (\ref{GG}). Moreover, the equation for $\ddot H$ is derived as
\begin{equation}
\frac{6H\ddot H}{\kappa^2}-2304\ddot H H^7 f_{GG}\simeq 2V'(\phi)\ddot\phi
-2HV'(\phi)\dot\phi\,\epsilon\,,\quad\epsilon=-\frac{\dot H}{H^2}\,.\label{47}
\end{equation}
Let us introduce the quadratic potential (\ref{pot1}) and assume the following form for $f(G)$,
\begin{equation}
f(G)=\gamma G^n\,,\quad n\neq 1\,,\label{AnG}
\end{equation}
where $\gamma$ is a dimensional constant such that
$[\gamma]=[M_{Pl}^{4(1-n)}]$,
and $n$ is a number. For $n=1$ we recover Einstein's gravity being the Gauss-Bonnet a topological invariant in four dimension.\\ 
\\
The simplest non trivial case of (\ref{AnG}) is given by $n=1/2$, for which the dimension of $\gamma$ is $[\gamma]=[M_{Pl}^2]$. We may assume $0<\gamma$ and
 introduce an effective mass of the theory as
\begin{equation}
m_\text{eff}=m\sqrt{\frac{3 M_{Pl}^2}{3M_{Pl}^2+8\pi\sqrt{6}\gamma}}\,,
\quad 0<\gamma\,,m_\text{eff}<m\,,
\end{equation}
such that, in analogy with (\ref{38})--(\ref{39}), one has the following solution of (\ref{45}),
\begin{equation}
H_\text{dS}=-2\sqrt{\frac{\pi}{3}}\frac{m_\text{eff}\phi_\text{i}}{M_\text{Pl}}\,,\quad
\phi\simeq\phi_\text{i}+\frac{m_\text{eff} M_{Pl}}{\sqrt{12\pi}}(t-t_\text{i})\,,
\end{equation}
and
\begin{equation}
\frac{M_{Pl}}{\sqrt{12\pi}}<|\phi_\text{i}|<\frac{M_{Pl}^2}{2m_\text{eff}}\sqrt{\frac{3}{\pi}}
\simeq 0.5\frac{M_{Pl}^2}{m_\text{eff}}\,.\label{PlancklimitGB}
\end{equation}
The slow-roll parameters (\ref{slowrollpar}) and the $e$-folds (\ref{Nfold}) 
follow from (\ref{46})--(\ref{47}) as
\begin{equation}
\epsilon\simeq\frac{M_{Pl}^2}{4\pi\phi_\text{i}^2}\left(\frac{m}{m_\text{eff}}\right)^2\,,\quad
\eta\simeq\frac{M_{Pl}^2}{4\pi\phi_\text{i}^2}\left(\frac{m}{m_\text{eff}}\right)^2\,,
\quad
\mathcal N\simeq\frac{2\pi\phi_\text{i}^2}{M_{Pl}^2}\left(\frac{m_\text{eff}}{m}\right)^2\,,
\end{equation}
and for $m_\text{eff}=m$ we recover (\ref{40}).
Thus, for large boundary values of the field $\phi_{i}$, the $e$-folds $\mathcal N$ can be large enough and the slow-roll parameters are very small during inflation, since
\begin{equation}
\epsilon\simeq\eta\simeq\frac{1}{2\mathcal N}\,,
\end{equation}
like in the case of $f(G)=0$. 

The amplitude of the primordial scalar power spectrum (\ref{spectrum}) of the model is
\begin{equation}
\Delta_{\mathcal R}^2=\frac{16}{3}\frac{m_\text{eff}^4\pi\phi^4}{M_{Pl}^6 m^2}\,.\label{add1}
\end{equation}
The spectral index and the tensor-to-scalar ratio (\ref{index}) read
\begin{equation}
n_s\simeq 1-\frac{M_{Pl}^2}{\pi\phi_\text{i}^2}\left(\frac{m}{m_\text{eff}}\right)^2\,,\quad
r=\frac{4M_{Pl}^2}{\pi\phi_\text{i}^2}\left(\frac{m}{m_\text{eff}}\right)^2\,,
\end{equation}
and in order to satisfy the Planck data (see under Eq.~(\ref{index})) one must require
\begin{equation}
0.0324<
\frac{M_{Pl}^2}{\pi\phi_\text{i}^2}\left(\frac{m}{m_\text{eff}}\right)^2
<0.0470\,,
\end{equation}
or, in terms of the $e$-folds number,
\begin{equation}
42<\mathcal N<61\,.\label{Ncase1}
\end{equation}
In the range $55<\mathcal N<61$ inflation is considered viable (we also must stress that acceleration continues after the slow-roll approxiamation and ends only when $\epsilon=1$).
Thus, 
\begin{equation}
2.6\frac{m M_{Pl}}{m_\text{eff}}<|\phi_i|<3.1\frac{m M_{Pl}}{m_\text{eff}}\,.\label{add2}
\end{equation}
Despite to the fact that the spectral index $n_s$ is viable, the model presents the same criticism of the massive scalar field in General Relativity framework concerning the tensor-to-scalar ratio $r$, which results to be $r\sim 0.13$, slightly bigger than the Planck constrain at $r<0.11$.
We note that, since the curvature during inflation cannot exceed the Planck mass, the condition (\ref{add2}) with (\ref{PlancklimitGB}) leads to
\begin{equation}
m<0.2 M_{Pl}\,,
\end{equation}
and in order to recover the primordial scalar power spectrum (\ref{add1}) $\Delta_\mathcal R^2\simeq 10^{-9}$, it must be $m\simeq 10^{-6} M_{Pl}$.
The Gauss-Bonnet contribution $f(G)=\gamma\sqrt{G}$ to the action is compatible with the inflationary scenario from scalar field with quadratic potential, but leads to a tensor-to-scalar ratio slightly bigger than the one given by the Planck data. If $0<\gamma$, the inflation is realized at curvature smaller respect to the classical case with $\gamma=0$ (on the other side, if $-3M_{Pl}^2/(8\pi\sqrt{6})<\gamma<0$, we obtain the opposite behaviour), but a suitable setting of the initial value of the field permits to recover the same spectral index.\\
\\
Let us take now $1<n$ in (\ref{AnG}): in this case, the value of $f(G)\sim \gamma R^{2n}$ is dominant respect to the Hilbert Einstein term during inflation when
\begin{equation}
\left(\frac{M_{Pl}^2}{|\gamma|}\right)^{\frac{1}{2n-1}}<R<M_{Pl}^2\,,\label{condR}
\end{equation}
such that the equations (\ref{45})--(\ref{47}) can be solved by neglecting the contribution coming from $R/\kappa^2$.
The de Sitter solution exists and it is real if $\gamma<0$,
\begin{equation}
H_\text{dS}\simeq\Phi\left(-\phi\right)^{1/2n}\,,\quad
\phi\simeq\phi_\text{i}+(-\phi_\text{i})^{\frac{2n-1}{2n}}\frac{m^2}{3\Phi }(t-t_\text{i})\,,
\quad
\Phi=\left(\frac{m}{\sqrt{2(24)^n\gamma(1-n)}}\right)^{\frac{1}{2n}}\,,
\end{equation}
with $[\Phi]=[M_{Pl}]^{\frac{2n-1}{2n}}$.
The slow-roll parameters (\ref{slowrollpar}) and the $e$-folds number (\ref{Nfold}) read
\begin{equation}
\epsilon\simeq\frac{m^2}{6n\Phi^2(-\phi_\text{i})^\frac{1}{n}}\,,\quad
\eta\simeq\frac{m^2}{3\Phi^2(-\phi_\text{i})^\frac{1}{n}}\,,\quad
\mathcal N\simeq\frac{3n(-\phi_\text{i})^\frac{1}{n}\Phi^2}{m^2}\,.
\end{equation}
For large values of $\phi_\text{i}$ the $e$-folds is large and the slow-roll parameters small. However, we will see that the Planck constraints limit the magnitude of $\phi_\text{i}$.
We observe
\begin{equation}
\epsilon\simeq\frac{1}{2\mathcal N}\,,\quad\eta\simeq\frac{n}{\mathcal N}\,,
\end{equation}
such that $\epsilon<\eta$.
The amplitude of the primordial scalar power spectrum (\ref{spectrum}) is given by
\begin{equation}
\Delta_{\mathcal R}^2=\frac{6n(-\phi_\text{i})^{2/n}\Phi^4}{m^2 M_{Pl}^2\pi}\,,
\end{equation}
and the spectral index and the tensor-to-scalar ratio (\ref{index}) are derived as
\begin{equation}
n_s\simeq 1-\frac{(3-2n)m^2}{3n\Phi^2(-\phi_\text{i})^{\frac{1}{n}}}\,,\quad
r=\frac{8 m^2}{3n\Phi^2(-\phi_\text{i})^{\frac{1}{n}}}\,.
\end{equation}
In order to satisfy the Planck data we must require
\begin{equation}
0.0324<
\frac{(3-2n)m^2}{3n\Phi^2(-\phi_\text{i})^{\frac{1}{n}}}
<0.0470\,,\quad 1<n<\frac{3}{2}\,.\label{condizionen}
\end{equation}
The condition on $n$ is quite restrictive and imply that inflation described by the model takes place very near to the Planck scale. For example, if $n\simeq 3/2^-$, it follows from (\ref{condR}),
\begin{equation}
\frac{M_{Pl}}{\sqrt{|\gamma|}}<R<M_{Pl}^2\,,
\end{equation}
but in this case the magnitude of the boundary value of the field in (\ref{condizionen}) cannot be very large and the $e$-folds of the model is quite small. A better fit of the cosmological data may be found for a value of $n$ between one and $3/2$, but the inflationary scenario produced by the model cannot be defined ``chaotic'' due to the restrictions on the boundary of the field. 

Viable inflation based on the account of $\gamma G^n\,,\gamma<0\,,1<n$, could be realized only if $1<n<3/2$, but, as soon as $n$ is close to $3/2$, the magnitude of the field is small even if the curvature is extremely near to the Planck scale. On the other hand, if $n$ is close to one, condition (\ref{condR}) is not well satisfied and the model turns out to be the one with $f(G)=0$. In the last part of the next subsection, we will reconsider such a model by adding a contribution from $R^n$: we will see that also in this case the conditions on $n$ for a feasible inflation do not change.

\subsection{$f(R^2,P, Q)$ power-law models \label{RPQ1}}

In this subsection, we will consider an explicit model of $f(R^2,P,Q)$ in the context of chaotic inflation from quadratic potential. To simplify the problem, we will rewrite $P, Q$ as funcions of the square of the Ricci scalar $R^2$, the Gauss-Bonnet invariant $G$ and the square of the Weyl tensor $C^2$ as in (\ref{superrelation}). Since the contribution of the Weyl tensor is identically null on FRW metric, we can reduce the theory to $f(R^2, G)$-gravity, and the equation (\ref{N}) with (\ref{GG}) reads
\begin{equation}
\frac{3H^2}{\kappa^2}+\left(f-R f_R-G f_G\right)+6H^2 f_R+6H \dot f_R+24H^3\dot f_G=
\frac{\dot\phi^2}{2}+V(\phi)\,.\label{equazioneR2G}
\end{equation}
Thus, the de Sitter solution is derived from
\begin{eqnarray}
&&\frac{3H_\text{dS}^2}{\kappa^2}+\left(f(R_\text{dS}^2, G_\text{dS})-R_\text{dS} f_R(R_\text{dS}^2, G_\text{dS})-G_\text{dS} f_G(R_\text{dS}^2, G_\text{dS})\right)+6H_\text{dS}^2 f_R(R_\text{dS}^2, G_\text{dS})=
V(\phi)\,,\nonumber\\
&&\hspace{1cm}R_\text{dS}=12 H_\text{dS}^2\,,\quad G_\text{dS}=24 H_\text{dS}^4\,.\label{dSR2G}
\end{eqnarray}
The derivative of (\ref{equazioneR2G}) in the slow-roll approximation (\ref{slowrollcondition}),
\begin{equation}
\frac{6H\dot H}{\kappa^2}+12H\dot H f_R+144H^3\dot H f_{RR}-1152H^5\dot H f_{RG}-2304\dot H H^7 f_{GG}\simeq V'(\phi)\dot\phi\,,
\end{equation}
corresponds to (\ref{Hdotgeneric}) with (\ref{GG}), and the equation for $\ddot H$ is given by
\begin{equation}
\hspace{-1cm}\frac{6H\ddot H}{\kappa^2}+12H\ddot H f_R+144H^3\ddot H f_{RR}-1152H^5\ddot H f_{RG}-2304\ddot H H^7 f_{GG}\simeq 2V'(\phi)\ddot\phi-2HV'(\phi)\dot\phi\,\epsilon\,,\quad\epsilon=-\frac{\dot H}{H^2}\,.
\end{equation}
For our purpose, let us take the following Ansatz for $f(R^2, P,Q)$-model,
\begin{equation}
f(R^2,P,Q)=\alpha R^{2n}+\beta P^m+\gamma Q^p\,,\label{explRPQ}
\end{equation}
where $\alpha, \beta, \gamma$ are dimensional constant and $n\,,m\,,p$ numbers. We get from (\ref{superrelation}),
\begin{equation}
f(R^2, G, C^2)=\alpha R^{2n}+\beta\left(\frac{C^2}{2}-\frac{G}{2}+\frac{R^2}{3}\right)^m
+\gamma\left(2C^2-G+\frac{R^2}{3}\right)^p\,.
\end{equation}
The behaviour of this model on FRW space-time correponds to the behaviour of
\begin{equation}
f(R^2, G)=\alpha R^{2n}+\beta\left(\frac{R^2}{3}-\frac{G}{2}\right)^m
+\gamma\left(\frac{R^2}{3}-G\right)^p\,,\label{RPQGC}
\end{equation}
being the Weyl tensor identically zero on FRW metric.
To study how inflation can be realized from such a kind of theory, we must do some assumption.
For $m=p=1$, the model reduces to $f(R^2)$-gravity, being $G$ a topological invariant in four dimension (its contribution in (\ref{equazioneR2G}) drops down), and we get
\begin{equation}
f(R^2)=\alpha R^{2n}+\xi R^{2}\,,\quad\xi=\frac{\beta+\gamma}{3}\,,\quad m=p=1\,.
\end{equation}
For $n\leq 1$, we deal in fact with a $R^2$ correction to standard gravity. In this case, at subplanckian scale, the Hilbert Einstein term $R/\kappa^2$ is dominant in the action, and inflation has a corresponding (viable) description in the so-called Einstein frame~\cite{Staro} after a conformal transformation of the metric. In literature we have many studies about inflation from $R^2$ in Einstein frame~\cite{Sta1}, or inflation from $R^2$ combined with other curvature invariants coming from trace-anomaly, quantum corrections or string inspired theories~\cite{q3, Sta2}. For $1<n$ we obtain more general power-low corrections to General Relativity~\cite{Sta3}: note that if we neglect the Einstein's term we do not have real de Sitter solution for positive values of the potential, and $n$ must remain close to one.\\
\\ 
One simple non trivial case is given by $n=m=p=2$ in (\ref{explRPQ})--(\ref{RPQGC}), for which we get
\begin{equation}
f(R^2, G)=\tilde\alpha R^{4}+\tilde\beta G^2+\tilde\gamma R^2G\,,\quad\xi=\left(\alpha-\frac{\gamma}{9}\right)\quad n=m=p=2\,,\label{ppmm}
\end{equation}
where
\begin{equation}
\tilde\alpha=\alpha+\frac{\beta}{9}+\frac{\gamma}{9}\,,\quad
\tilde\beta=\frac{\beta}{4}+\gamma\,,\quad
\tilde\gamma=-\frac{\beta}{3}-\frac{2\gamma}{3}\,,\label{ppmm2}
\end{equation}
with $[\alpha]=[\beta]=[\gamma]=[1/M_{Pl}^4]$.
Inflation takes place in high curvature limit, 
\begin{equation}
\left(\frac{M_{Pl}^2}{\delta}\right)^\frac{1}{3}<R<M_{Pl}^2\,,\label{rangeRPQ}
\end{equation}
where $\delta$ is a term with the dimension and magnitude of $\tilde\alpha\,,\tilde\beta\,,\tilde\gamma$: in this case
the Hilbert-Einstein contribution can be neglected in the action.
The de Sitter solution is derived from (\ref{dSR2G}) as
\begin{equation}
H_\text{dS}=\Phi(-\phi_\text{i})^{1/4}\,,\quad
\phi\simeq\phi_\text{i}+(-\phi_\text{i})^{\frac{3}{4}}\frac{m^2}{3\Phi }(t-t_\text{i})
\,,\quad\Phi=\frac{(m)^{1/4}}{\left(2^{7/8}3^{1/4}\right)\left(-36\tilde\alpha-\tilde\beta-6\tilde\gamma\right)^{1/8}}\,,
\label{dSmodellino}
\end{equation}
with $[\Phi]=[M_{Pl}^{3/4}]$ and $(36\tilde\alpha+\tilde\beta+6\tilde\gamma)<0$.
The behaviour of the field, as usually, is governed by (\ref{cons2}).
The slow roll parameters and the $e$-folds number read
\begin{equation}
\epsilon\simeq\frac{m^2(36\tilde\alpha+\tilde\beta+6\tilde\gamma)}{12(\tilde\beta+3\tilde\gamma-72\tilde\alpha)\sqrt{-\phi_\text{i}}\Phi^2}\,,\quad
\eta\simeq\frac{m^2}{3\sqrt{-\phi_\text{i}}\Phi^2}\,,\quad
\mathcal N\simeq\frac{6\sqrt{-\phi_\text{i}}\Phi^2}{m^2}\,.\label{mio}
\end{equation}
In order to have $\epsilon>0$ with a real solution for the de Sitter solution (\ref{dSmodellino}), it must be
\begin{equation}
\frac{\tilde\beta+3\tilde\gamma}{72}<\tilde\alpha<-\frac{(\tilde\beta+6\tilde\gamma)}{36}\,.
\label{cond11}
\end{equation}
From (\ref{mio}) we get
\begin{equation}
\epsilon\simeq\frac{(36\tilde\alpha+\tilde\beta+6\tilde\gamma)}{2(\tilde\beta+3\tilde\gamma-72\tilde\alpha)\mathcal N}\,,\quad
\eta\simeq\frac{2}{\mathcal N}\,.
\end{equation}
The amplitude of the primordial scalar power spectrum (\ref{spectrum}) is derived as
\begin{equation}
\Delta_{\mathcal R}^2=\frac{12(\tilde\beta+3\tilde\gamma-72\tilde\alpha)(-\phi_\text{i})\Phi^4}{m^2 M_{Pl}^2\pi(36\tilde\alpha+\tilde\beta+6\tilde\gamma)}\,,
\end{equation}
and for the spectral index and the tensor-to-scalar ratio (\ref{index}) one finds
\begin{equation}
n_s\simeq 1-\frac{m^2(396\tilde\alpha-\tilde\beta+6\tilde\gamma)}{6(\tilde\beta+3\tilde\gamma-72\tilde\alpha)\sqrt{-\phi_\text{i}}\Phi^2}\,,\quad
r=\frac{4m^2(36\tilde\alpha+\tilde\beta+6\tilde\gamma)}{3(\tilde\beta+3\tilde\gamma-72\tilde\alpha)\sqrt{-\phi_\text{i}}\Phi^2}\,.
\end{equation}
To satisfy the Planck data it must be required
\begin{equation}
0.0324<
\frac{m^2(396\tilde\alpha-\tilde\beta+6\tilde\gamma)}{6(\tilde\beta+3\tilde\gamma-72\tilde\alpha)\sqrt{-\phi_\text{i}}\Phi^2}
<0.0470\,,\label{ultult}
\end{equation}
but also in this case the tensor-to-scalar ratio $r$ is bigger than the Planck contraints, being on the order $r\sim 0.26$. In order to recover $\Delta_\mathcal R^2\sim 10^{-9}$, it is enough to have $m\sim 10^{-6} M_{Pl}$.
We immediatly see from (\ref{cond11}),
\begin{equation}
\tilde\alpha<\frac{\tilde\beta-6\tilde\gamma}{396}\,.\label{cond22}
\end{equation}
Conditions (\ref{cond11}) and (\ref{cond22}) must be satisfied simultaneously. We have several possibilities. If $\tilde\gamma=0$, namely $\beta=-2\gamma$ in (\ref{explRPQ})--(\ref{RPQGC}),
\begin{equation}
\tilde\beta<0\,,\quad\frac{\tilde\beta}{72}<\alpha<\frac{\tilde\beta}{396}\,,\quad\tilde\gamma=0\,,
\end{equation}
but in this case the theory is affected by antigravitational effects during inflation, since 
the effective gravitational constant of the model, $G_\text{eff}=G_N/(1+2\kappa^2 f_R(R^2, G))$, $G_N$ being the Newton constant, results to be negative if $\tilde\alpha<0$. 

In general, if $\tilde\gamma<0$ and $6\tilde\gamma<\tilde\beta<-6\tilde\gamma$, conditions (\ref{cond11}) and (\ref{cond22}) can be satisfied for positive values of $\tilde\alpha$ with the possibility to avoid antigravitational effects. By using $\mathcal{N}$ of (\ref{mio}) in (\ref{ultult}) we get
\begin{equation}
\frac{21(\tilde\beta+3\tilde\gamma-72\tilde\alpha)}{(396\tilde\alpha-\tilde\beta+6\tilde\gamma)}<\mathcal N<
\frac{31(\tilde\beta+3\tilde\gamma-72\tilde\alpha)}{(396\tilde\alpha-\tilde\beta+6\tilde\gamma)}\,,
\end{equation}
and to have $55<\mathcal N$ we must require $(\tilde\beta+3\tilde\gamma-72\tilde\alpha)\simeq 2(396\tilde\alpha-\tilde\beta+6\tilde\gamma)$, such that a suitable value of $\phi_\text{i}$ can reproduce a sufficient amount of inflation according with Planck results.

We have seen that the model $f(R^2, P, Q)=\alpha R^4+\beta P^2+\gamma Q^2$ with a massive scalar field may bring to a viable early-time acceleration with curvature near to the Planck scale, but the tensor-to-scalar ratio is not compatible with the Planck data (but, for example, may be compatible with the BICEP2 results~\cite{BICEP}). When the curvature decreases, we are out of the range (\ref{rangeRPQ}) and the Hilber-Einsten term becomes dominant in the action: at this point, the reahting processes for particle production take places and Friedmann expansion starts.\\
\\
To conclude this chapter, we would like to reconsider the model (\ref{AnG}) with $1<n$ of the previous subsection togeter with a power law function of the Ricci scalar in the context of $f(R^2, G)$-gravity, namely
\begin{equation}
f(R^2, G)=\alpha R^{2n}+\gamma G^n\,,\quad 1<n\,,
\end{equation} 
where $\alpha,\gamma$ are constants whose dimensions are $[\alpha]=[\gamma]=[1/M_{Pl}^{4(n-1)}]$. Inflation starts at high curvature, 
\begin{equation}
\left(\frac{M_{Pl}^2}{|\delta|}\right)^\frac{1}{2n-1}<R<M_{Pl}^2\,,
\end{equation}
where $\delta$ is a term with the dimension and the magnitude of $\alpha\,,\gamma$, such that the Hilbert Einstein contribution $R/\kappa^2$ can be ignored into the action. 
In the presence of massive scalar field, the de Sitter solution of the model reads
\begin{equation}
H_\text{dS}=\Phi(-\phi_\text{i})^{1/(2n)}\,,\quad
\phi\simeq\phi_\text{i}+(-\phi_\text{i})^{\frac{2n-1}{2n}}\frac{m^2}{3\Phi }(t-t_\text{i})\,,
\quad\Phi=
\frac{m^{1/2n}}{\left(2^{1+3n} 3^n (n-1)(-6^n\alpha-\gamma)\right)^{1/4n}}\,,
\end{equation}
with $[\Phi]=[M_{Pl}^{(2n-1)/2n}]$ and $(6^n\alpha+\gamma)<0$.
Therefore, the slow roll parameters and the $e$-folds number are derived as
\begin{equation}
\epsilon\simeq\frac{m^2(n-1)(6^n\alpha+\gamma)}{6n((n-1)\gamma-6^n n\alpha)(-\phi_\text{i})^{1/n}\Phi^2}\,,\quad
\eta\simeq\frac{m^2}{3(-\phi_\text{i})^{1/n}\Phi^2}\,,\quad
\mathcal N\simeq\frac{3n(-\phi_\text{i})^{1/n}\Phi^2}{m^2}\,.
\end{equation}
In order to get $\epsilon>0$ with a real solution for the de Sitter solution we must find
\begin{equation}
\frac{(n-1)\gamma}{6^n n}<\alpha<-\frac{\gamma}{6^n}\,.\label{cond111}
\end{equation}
We also observe
\begin{equation}
\epsilon\simeq\frac{(n-1)(6^n\alpha+\gamma)}{2((n-1)\gamma-6^n n\alpha)\mathcal N}\,,\quad
\eta\simeq\frac{n}{\mathcal N}\,.
\end{equation}
Thus, the amplitude of the primordial scalar power spectrum (\ref{spectrum}) is derived as
\begin{equation}
\Delta_{\mathcal R}^2=\frac{6n(\gamma(n-1)-6^n n\alpha)(-\phi_\text{i})^{2/n}\Phi^4}{m^2 M_{Pl}^2\pi(n-1)(6^n\alpha+\gamma)}\,.
\end{equation}
and for the spectral index and the tensor-to-scalar ratio we get
\begin{equation}
n_s\simeq 1-
\frac{m^2(6^n(n(3+2n)-3)\alpha+(n(5-2n)-3)\gamma)}{3n((n-1)\gamma-6^n n\alpha)(-\phi_\text{i})^{1/n}\Phi^2}
\,,\quad
r=\frac{8 m^2(n-1)(6^n\alpha+\gamma)}{3n((n-1)\gamma-6^n n\alpha)(-\phi_\text{i})^{1/n}\Phi^2}\,.
\end{equation}
As a consequence, to reproduce the Planck data we must require
\begin{equation}
0.0324<
\frac{m^2(6^n(n(3+2n)-3)\alpha+(n(5-2n)-3)\gamma)}{3n((n-1)\gamma-6^n n\alpha)(-\phi_\text{i})^{1/n}\Phi^2}
<0.0470\,,
\end{equation}
and finally 
\begin{equation}
\alpha<\frac{(3-(5-2n)n)\gamma}{6^n(n(3+2n)-3)}\,.\label{cond222}
\end{equation}
In order to satisfy conditions (\ref{cond111}) and (\ref{cond222}), we can take $\gamma<0$ and $0<\alpha$ (avoiding antigravitational effects) only if
\begin{equation}
1<n<\frac{3}{2}\,,
\end{equation}
recovering the same result of (\ref{condizionen}) where the $R^{2n}$ contribution was not considered. In other words, the addition of a $R^{2n}$ contribution to the model in (\ref{AnG}) does not change the range of $n$, which remains quite close to one to reproduce the Planck results.

\section{Quartic potential}

Let us consider now a quartic field potential in the general action (\ref{action}), 
\begin{equation}
V(\phi)=\frac{\lambda\phi^4}{4}\,,\quad 0<\lambda\,,\label{pot2}
\end{equation}
where $\lambda$ is a positive adimensional constant.
In Einstein's gravity where $f(R^2,P,Q)=0$, for large and negative value of the field we get from (\ref{equazioneds}) and (\ref{cons2}),
\begin{equation}
H_\text{dS}=\sqrt{\frac{2\pi}{3}}\frac{\sqrt{\lambda}\phi_\text{i}^2}{M_\text{Pl}}\,,\quad
\phi\simeq\phi_\text{i}-\phi_\text{i}\frac{\sqrt{\lambda} M_{Pl}}{\sqrt{6\pi}}(t-t_\text{i})\,,\label{883}
\end{equation}
where, as usually, $\phi_\text{i}<0$ is the boundary value of inflation.
Thus, the slow-roll approximation (\ref{phiV}) is valid as soon as
\begin{equation}
\frac{M_{Pl}}{\sqrt{3\pi}}<|\phi_\text{i}|<M_{Pl}\left(\frac{3}{2\pi\lambda}\right)^{1/4}\,,\label{388}
\end{equation}
where $M_{Pl}/(\lambda)^{1/4}\ll 1$ and
the field may be larger than the Planck Mass during inflation. 
In this case, the $e$-folds and the slow-roll parameters (\ref{ciao}) read
\begin{equation}
\epsilon\simeq\frac{M_{Pl}^2}{\pi\phi_\text{i}^2}\,,\quad 
\eta\simeq\frac{3 M_{Pl}^2}{2\pi \phi_\text{i}^2}\,,\quad
\mathcal N\simeq \frac{\pi\phi_\text{i}^2}{M_{Pl}^2}\,,\label{104}
\end{equation}
and, for large $e$-folds, the slow-roll parameters 
are small and the spectral index $n_s$ in (\ref{index}) can satisfy the Planck data, with a larger value of the tensor-to-scalar ratio $r$.

In the following subsections, we will see some significative examples of $f(R^2, P, Q)$-gravity where chaotic inflation from field with quartic potential could be realized.

\subsection{Gauss-Bonnet models\label{GB2}}

As a first example, we will consider the Gauss-Bonnet model
\begin{equation}
f(G)=\gamma\sqrt{G}\,,\quad 0<\gamma,
\end{equation} 
$\gamma$ being a positive dimensional constant such that $[\gamma]=[M_{Pl}^2]$. As we have already seen in \S~\ref{GB1} this kind of correction to Einstein's gravity leads to a viable inflation in the presence of massive scalar field.

If we introduce the effective $\lambda_\text{eff}$ parameter,
\begin{equation}
\lambda_\text{eff}=\frac{\lambda M_{Pl}^2}{M_{Pl}^2+(4\pi/3)\gamma\sqrt{24}}\,,\quad
0<\gamma\,,\lambda_\text{eff}<\lambda\,,
\end{equation}
for the potential (\ref{pot2}), Equation~(\ref{45}) leads, on the de Sitter solution,
\begin{equation}
H_\text{dS}=\sqrt{\frac{2\pi}{3}}\frac{\sqrt{\lambda_\text{eff}}\phi_\text{i}^2}{M_\text{Pl}}\,,\quad
\phi\simeq\phi_\text{i}-\phi_\text{i}\frac{\sqrt{\lambda_\text{eff}} M_{Pl}}{\sqrt{6\pi}}(t-t_\text{i})\,,
\end{equation}
with
\begin{equation}
\frac{M_{Pl}}{\sqrt{3\pi}}<|\phi_\text{i}|<M_{Pl}\left(\frac{3}{2\pi\lambda_\text{eff}}\right)^{1/4}\simeq \frac{1.5 M_{Pl}}{\lambda_\text{eff}^{1/4}} \,,
\label{franco}
\end{equation}
in analogy with (\ref{883})--(\ref{388}).

The slow-roll parameters (\ref{slowrollpar}) and the $e$-folds (\ref{Nfold}) 
are derived from (\ref{46})--(\ref{47}) and result to be
\begin{equation}
\epsilon\simeq\frac{M_{Pl}^2}{\pi\phi_\text{i}^2}\left(\frac{\lambda}{\lambda_\text{eff}}\right)\,,\quad
\eta\simeq\frac{3 M_{Pl}^2}{2\pi\phi_\text{i}^2}\left(\frac{\lambda}{\lambda_\text{eff}}\right)\,,
\quad
\mathcal N\simeq\frac{\pi\phi_\text{i}^2}{M_{Pl}^2}\left(\frac{\lambda_\text{eff}}{\lambda}\right)\,,
\end{equation}
and for $\lambda_\text{eff}=\lambda$ we recover (\ref{104}). Since
\begin{equation}
\epsilon\simeq\frac{1}{\mathcal N}\,,\quad\eta\simeq\frac{3}{2\mathcal N}\,,
\end{equation}
we see that
for large boundary values of the field $\phi_{i}$, the $e$-folds $\mathcal N$ can be large enough and the slow-roll parameters very small during inflation. 

The amplitude of the primordial scalar power spectrum (\ref{spectrum}) of the model is given by
\begin{equation}
\Delta_{\mathcal R}^2=\frac{2\pi\lambda_\text{eff}^2\phi_\text{i}^6}{3M_{Pl}^6\lambda}\,,
\end{equation}
while the spectral index and the tensor-to-scalar ratio (\ref{index}) read
\begin{equation}
n_s\simeq 1-\frac{3M_{Pl}^2}{\pi\phi_\text{i}^2}\left(\frac{\lambda}{\lambda_\text{eff}}\right)\,,\quad
r=\frac{16M_{Pl}^2}{\pi\phi_\text{i}^2}\left(\frac{\lambda}{\lambda_\text{eff}}\right)\,.
\end{equation}
Thus, in order to satisfy the Planck data we must find
\begin{equation}
0.0324<
\frac{3M_{Pl}^2}{\pi\phi_\text{i}^2}\left(\frac{\lambda}{\lambda_\text{eff}}\right)
<0.0470\,,
\end{equation}
or, in terms of the $e$-folds number,
\begin{equation}
64<\mathcal N<93\,.\label{Ncase2}
\end{equation}
As a consequence, the boundary value of the field must be
\begin{equation}
4.5 M_{Pl}\sqrt{\frac{\lambda}{\lambda_\text{eff}}}<|\phi_i|<5.4 M_{Pl}\sqrt{\frac{\lambda}{\lambda_\text{eff}}}\,.
\end{equation}
As in the case of scalar field with quartic potential in the framework of General Relativity, the tensor-to-scalar ratio $r$ is bigger than the Planck bound ($r\sim 0.16$).
Since the curvature during inflation cannot exceed the Planck mass, the condition above with (\ref{franco}) leads to
\begin{equation}
\sqrt{\frac{\lambda}{\sqrt{\lambda_\text{eff}}}}<0.33\,,
\end{equation}
and in order to obtain $\Delta_\mathcal R^2\simeq 10^{-9}$, one must have 
$\sqrt{\lambda/\sqrt{\lambda_\text{eff}}}\sim 10^{-7}$.
In conclusion, the model $f(G)=\gamma\sqrt{G}\,,0<\gamma$, in the presence of scalar field with quadratic (see \S~\ref{GB1}) or quartic potential may lead to a viable inflationary scenario, but the tensor-to-scalar ratio $r$ exceed the Planck result. 
Moreover, in the case investigated in this subsection the $e$-folds in (\ref{Ncase2}) is larger than the $e$-folds in (\ref{Ncase1}). As a consequence, the slow-roll parameters of the model are smaller in the presence of scalar field with quartic potential respect to the quadratic potential case.

\subsection{$f(R^2, G)$ power-law models}

In this last subsection we will discuss a general form of $f(R^2, G)$ power-law model and we will see that, in the presence of scalar field with quartic potential, differently to the cases of quadratic potential analyzed in \S~\ref{RPQ1}, inflation cannot be realized. We also remember that, as we have already discussed, a $f(R^2, G)$-model can be seen as a representation of a $f(R^2, P, Q)$-theory on FRW space-time if we use (\ref{superrelation}) and therefore consider null the Weyl tensor and its contributes to the action.

Let us start by the folliwing $f(R^2, G)$-model,
\begin{equation}
f(R^2, G)=\alpha R^{2n}+\beta G^{n}+\gamma (R^2 G)^{n/2}\,,\quad 1<n\,,\label{lastA}
\end{equation}
$\alpha, \beta, \gamma$ being dimensional constants such that $[\alpha]=[\beta]=[\gamma]=[M_{Pl}^{4(1-n)}]$.
This correction to Einstein's gravity is dominant respect to the Hilbert-Einstein term $R/\kappa^2$ in the action at high curvature when
\begin{equation}
\left(\frac{M_{Pl}^2}{|\delta|}\right)^{\frac{1}{2n-1}}<R<M_{Pl}^2\,,
\end{equation}
where $\delta$ is a term with the dimension and the magnitude of $\alpha, \beta, \gamma$.
The de Sitter solution of the model for the quartic potential (\ref{pot2}) follows from  (\ref{dSR2G}) as
\begin{equation}
H_\text{dS}=\Phi(-\phi_\text{i})^{1/n}\,,\quad\Phi=\frac{\lambda^{1/(4n)}}{2^{(3n+2)/(4n)}3^{1/4}(n-1)^{1/(4n)}\left(-6^n\alpha-\beta-6^{n/2}\gamma\right)^{1/(4n)}}\,,
\end{equation}
with $[\Phi]=M_{Pl}^{(n-1)/n}$ and $6^n\alpha+\beta+6^{n/2}\gamma<0$. Thus, the slow-roll parameters (\ref{slowrollpar}) and the $e$-folds (\ref{Nfold}) are derived as
\begin{equation}
\epsilon\simeq\frac{2(n-1)\left(6^n\alpha+\beta+6^{n/2}\gamma\right)\lambda
(-\phi_\text{i})^{\frac{2(n-1)}{n}}}{3n\left((2\beta+6^{n/2}\gamma)(n-1)-2^{n+1}3^n n\alpha\right)\Phi^2}
\,,\quad
\eta\simeq\frac{\lambda(-\phi_\text{i})^{\frac{2(n-1)}{n}}}{\Phi^2}
\,,\quad
\mathcal N\simeq\frac{3\Phi^2}{\left(\frac{2(n-1)}{n}\right)\lambda\phi^{\frac{2(n-1)}{n}}}\,.
\end{equation}
We immediatly see that, for $1<n$, large value of $\phi_\text{i}$ leads to large slow-roll parameters and small $e$-folds, rendering the inflationary scenario unrealistic.

This result cannot be considered a general behaviour of $f(R^2, P, Q)$- or $f(R^2, G)$-gravitational models: however, we would like to note that our Ansatz (\ref{lastA}) repersents a quite generic and reasonable power-law model of $f(R^2, G)$. For such a form of correction to Einstein gravity we can say that chaotic inflation from scalar field does not work in the presence of quartic potential.

\section{Conclusions}

In this paper, we have investigated chaotic inflation with scalar field subjected to potential in the framework of $f(R^2, P, Q)$-modified gravity, namely the gravitational action of the theory includes a correction based on an (arbitrary) function of the square of the Ricci scalar $R^2$ and the contractions of Ricci ($P$) and Riemann ($Q$) tensors. This form of modified gravity is quite general, and the curvature invariants under consideration may be related with quantum corrections to General Relativity or string inspired theories. To derive the Equations of motion on flat Friedmann-Robertson-Walker space-time we used a method based on Lagrangian multipliers and we treated the curvature invariants as independent functions: as a consequence, we deal with a system of second order differential equations simplifying the analysis of the model, which leads in principle to fourth order differential equations. We note that on FRW metric every $f(R^2, P, Q)$-theory can be reduced to a Gauss-Bonnet $f(R^2, G)$-theory, since in fact one of the curvature invariant can be expressed in terms of the other two. This feauture is manifest by replacing the contractions of Ricci and Riemann tensors $P, Q$ with the Gauss-Bonnet $G$ and the square of the Weyl tensor $C^2$: on FRW metric the Weyl tensor and its derivatives disappear and we can drop down its contribute from the Lagrangian. We used the $f(R^2, G)$-representation to study our models, since in this way  the Equations of motion result to be simplified. 

Chaotic inflation from scalar field with potential can be realized in the framework of higher derivative models as well as in the framework of General Relativity, but, despite to the fact that the continuity equation of the field keeps the same form in the two theories, the Hubble parameter and its derivatives depend on the field potential in different ways. Inflation must satisfy several constrains to be ``viable'': the (quasi) de Sitter solution must take place and the slow-roll approximations must be valid, it means, the slow-roll parameters must be small, the $e$-folds sufficiently large to guarantee the thermalization of observed universe, and the spectral index and the tensor-to-scalar ratio must satisfy the Planck data. We presented the general formalism to investigate inflation and its characteristic paramters in $f(R^2, P, Q)$-gravity with scalar field and we furnished several explicit examples. In the specific, we investigated two well-known forms of chaotic inflation, namely chaotic inflation from massive scalar field (quadratic potential) and chaotic inflation from field with quartic potential. We confronted the results in the framework of General Relativity and in the one of our modified theory giving some examples of corrections to Einstein's gravity based on power-law functions of the Gauss- Bonnet or of the other curvature invariants under investigation. The (positive) corrections based on the square root of the Gauss-Bonnet are on the same order of magnitude of the Hilbert-Einstein term in the action at high curvature and permit to realize an inflationary scenario similar to the Einstein's case. More interseting are the higher power-law functions of the Gauss-Bonnet and the other curvature invariants. In this cases, at high curvature the modification to gravity is dominant with respect to the Hilbert-Einstein term in the action and drives inflation. Thus, by fitting the parameters and the boundary value of the field, we may recover a viable inflation in the case of scalar field with quadratic potential, but with quartic potential the inflationary scenario appears to be unrealistic. We stress that this result cannot be read as a general behaviour of $f(R^2, P, Q)$- or $f(R^2, G)$-gravitational theories, even if  our Ansatz for the presented power-law models is  quite generic and reasonable. Moreover, even in the presence of our Ansatz, we cannot state that this models are not able to reproduce an early -time acceleration in agreement with Planck data, but only that in the context of chaotic inflation induced by large magnitude value of the inflaton inflation is not viable.  

A last remark is in order about the tensor-to-scalar ratio number of the models investigated in the present work, which results to be larger than the Planck contrain. This feauture is quite general in chaotic inflation from power-law potential, but, since the exact value of the tensor-to-scalar ratio is still object of a debated question, we think that this kind of models have to be still investigated. On the other hand, the attempt of our study is to furnish a general formalism for chaotic inflation in  higher derivative gravity theories, which is valid independently on the specific examples here analyzed.

Other works on higher derivative corrections to Einstein's gravity, FRW $f(G)$-gravity  and inflation can be found in Refs.~\cite{altro1, altro2, altro3}.

\end{document}